\documentclass[usenatbib]{mnras}
\usepackage[dvips]{color}
\usepackage{ulem}
\usepackage{url}

\usepackage{graphicx}
\usepackage{lscape}

\def\arcmin{\hbox{$^\prime$}}
\def\arcsec{\hbox{$^{\prime\prime}$}}

\def\flux{erg s$^{-1}$ cm$^{-2}$}
\def\lum{erg s$^{-1}$}

\def\aap{A\&A}
\def\mnras{MNRAS}
\def\apjs{ApJS}

\def\apjl{ApJLett}

\def\2s{2S\,1553$-$542}

\title[2S\,1553$-$542: a pulsar on the far side of the Galaxy]
{\2s: a Be/X-ray binary pulsar on the far side of the Galaxy}
\author[Lutovinov et al.]{Alexander A. Lutovinov$^{1,2}$\thanks{E-mail: aal@iki.rssi.ru},
David A.H. Buckley$^{3}$,
Lee J. Townsend$^{4}$,
\newauthor Sergey S. Tsygankov$^{5,1}$ and Jamie Kennea$^{6}$
\\
$^1$ Space Research Institute, Profsoyuznaya str. 84/32, Moscow, 117997, Russia  \\
$^2$ Moscow Institute of Physics and Technology, Institutskiy per. 9, Dolgoprudny,
Moscow Region, 141700, Russia \\
$^3$ South African Astronomical Observatory, PO Box 9, Observatory Road, Observatory 7935, South Africa  \\
$^4$ Astrophysics, Cosmology and Gravity Centre, Department of Astronomy, University of Cape Town, Rondebosch 7701, South Africa\\
$^5$ Tuorla Observatory, Department of Physics and Astronomy, University of Turku, V\"ais\"al\"antie 20, FI-21500 Piikki\"o, Finland \\
$^6$ Department of Astronomy and Astrophysics, The Pennsylvania State University, University Park, PA 16802, USA}

\begin{document}

\date{Accepted .... Received ...}

\pagerange{\pageref{firstpage}--\pageref{lastpage}} \pubyear{2016}

\maketitle

\label{firstpage}

\begin{abstract}

{We report the results of a comprehensive analysis of X-ray ({\it Chandra}
and {\it Swift} observatories), optical (Southern African Large Telescope, {\it SALT})
and near-infrared (the {\it VVV} survey) observations of the Be/X-ray
binary pulsar \2s. Accurate coordinates for the X-ray source are determined and
are used to identify the faint optical/infrared counterpart for the first time.
Using {\it VVV} and SALTICAM photometry, we have constructed the spectral energy
distribution (SED) for this star and found a moderate NIR excess that is
typical for Be stars and arises due to the presence of circumstellar material (disk).
A comparison of the SED with those of known Be/X-ray binaries has
allowed us to estimate the spectral type of the companion star as B1-2V and
the distance to the system as $>15$ kpc. This distance estimation is supported
by the X-ray data and makes \2s\ one of the most distant X-ray binaries within
the Milky Way, residing on the far side in the Scutum-Centaurus arm or even further.}

\end{abstract}


\section{Introduction}

Determining the nature of Galactic X-ray sources through their optical
counterparts and measurements of their distance is key not only in the study
of individual sources, but also in population studies of different classes of
objects. In the case of X-ray emission from a compact object (namely an X-ray
binary) the former allows one to understand physical processes near the
compact object. Only a good knowledge of the optical counterparts to sources
of X-ray emission allows testing of different theoretical models of the
emission mechanisms and accretion processes. In turn, properties of different
populations of X-ray sources in our and other galaxies contain information
about fundamental physical mechanisms responsible for their formation and
evolution and about properties of the host galaxy itself.

Distance estimations can be made relatively straightforwardly for X-ray
sources in other galaxies, but is a non-trivial task in our own galaxy. There
are at least two problems to contend with: the high density of stars, which
complicates the process of optical identification, and the large and variable
interstellar absorption, which complicates measurements of different
properties of the source, including its distance.

These difficulties have led to the situation where a significant
number of Galactic X-ray sources are still unclassified or have no reliable
distance measurements \citep[][]{liu06,liu07}.

In this work we utilize high quality X-ray, optical and near-infrared (NIR)
data to localize and identify the poorly studied transient X-ray pulsar \2s.
It was discovered in 1975 by the {\it SAS-3} observatory
\citep{1976IAUC.2959....2W}. Later a strong coherent variability, with
a period of 9.3 s, was found \citep{1982IAUC.3667....3K} in the source
light curve. Subsequent observations of \2s revealed strong outbursts in
2007 \citep{2007ATel.1345....1K} and most recently in 2015
\citep{2015ATel.7018....1S}. Observations with the {\it NuSTAR} observatory
and {\it Fermi}/GBM monitor during the 2015 outburst led
\citet{tsygankov2016} to the discovery of a cyclotron absorption line in the
source spectrum at $\simeq23.5$ keV and to improve the accuracy of the binary
parameters. Moreover these authors also estimated the distance to the system
$d\simeq20$ kpc.

The possible nature of \2s\ as a pulsating Be/X-ray binary system
was first suggested on the basis on its transient activity
\citep{1983ApJ...274..765K} and X-ray spin modulation. However, no optical
or infrared counterpart had been directly determined until now.
This made \2s\ one of the first transient X-ray binaries to be
discovered with a probable Be-companion, though the confirmation of this
and the distance to the system were still to be confirmed.
Moreover, the X-ray behaviour of the source is quite unusual --
in particular, the source never displays type I outbursts. \citet{oka_neg01}
proposed the model of the truncated disc in which such a behaviour
is naturally explained for systems with a low eccentricity.
Indeed, current measurements of \citet{tsygankov2016} revealed a very low
eccentricity in this system $e\simeq0.035$.

Based on recent data from the {\it Chandra} and {\it Swift} observatories,
Southern African Large Telescope (SALT) and the {\it VVV} survey, we have
measured for the first time an accurate position for the X-ray pulsar \2s,
leading to the identification of its infrared counterpart and an
estimation of its distance.

\section{X-ray observations and source position}

\subsection{Observations and data reduction}

\begin{figure}
\centering
\includegraphics[width=0.98\columnwidth,bb=3 269 547 535,clip]{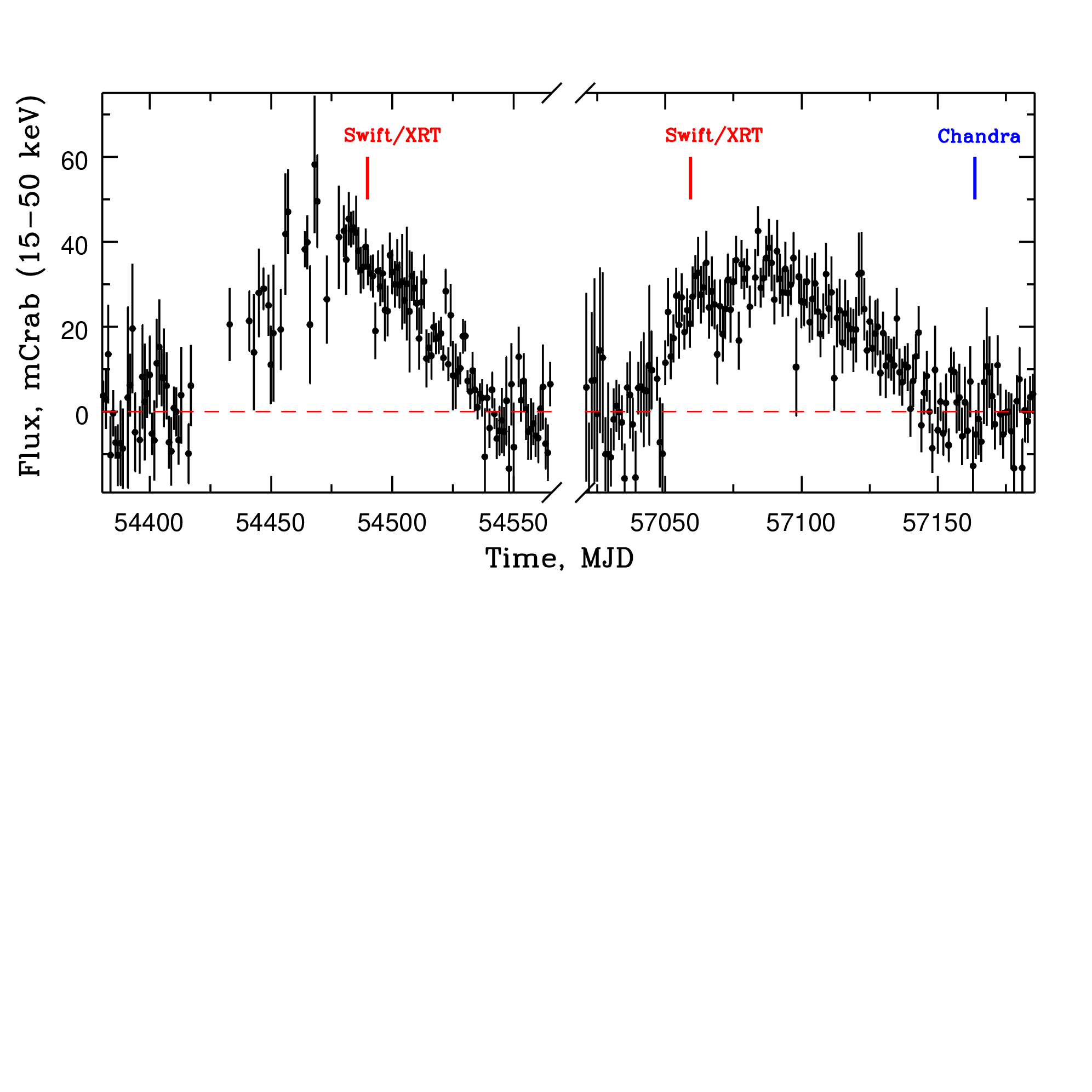}
\caption{One-day averaged light curve of the X-ray pulsar \2s\ obtained with the
{\it Swift}/BAT monitor in the 15-50 keV energy band for two outbursts: Dec 2007
-- Mar 2008 and  Jan--May 2015. Red marks correspond to the time of XRT
observations and the blue mark to the {\it Chandra} one. The period between
outbursts has been truncated for viewing purposes.}\label{lcurve}
\end{figure}

The source light curve obtained with the {\it Swift}/BAT telescope in the
$15-50$ keV energy band for two outbursts is shown in Fig.\ref{lcurve}. It is
seen that both outbursts are quite similar: the duration is about 3 months
and the maximum intensity around $40-50$ mCrab in the $15-50$ keV energy
band. The outburst light curves, duration and absence of orbital modulation
are all indications that they are type II outbursts.

Observations in soft X-rays were performed twice with the {\it Swift}/XRT
telescope (Obs.IDs 00031096001 and 00031096002, red lines in
Fig.\ref{lcurve}) and once with the ACIS instrument onboard the {\it Chandra}
observatory (ObsId. 17662, blue line in Fig.\ref{lcurve}). Both XRT
observations were carried out in the Photon Counting (PC) mode with a
relatively high source flux that led to pile-up in the data. It was taken
into account in the subsequent analysis according to the XRT Data Analysis
threads\footnote{http://www.swift.ac.uk/analysis/xrt/index.php}. The data
collected by the ACIS instrument with a total exposure of $\sim5$ ks were
reduced with the standard software package {\sc CIAO
4.7}\footnote{http://cxc.harvard.edu/ciao/} with CALDB v4.6.5. Note, that
these data did not suffer from pile-up.

\subsection{Refinement of the source coordinates}
\label{sec:xraypos}

As mentioned above, the optical companion of \2s\ is not yet
known, mainly because of the lack of an accurate localization of
the X-ray source. The {\it Swift}/XRT
telescope determined the position of \2s\ during the 2007-2008 outburst to be
(J2000): R.A.= 15$^{\rm h}$57$^{\rm m}$47.75$^{\rm s}$, Dec.=
-54$^\circ$24\arcmin52.4\arcsec\ with a 3.5 arc-second error
\citep{2008ATel.1371....1B}. This was a significant improvement on the
originally determined error circle of 35 arc-seconds
({\it SAS-3} aperture size), but nevertheless did not lead to the
determination of the optical counterpart.

To determine the position of \2s\ with a higher accuracy, we
triggered a TOO observation with the {\it Chandra} observatory on May
21, 2015 (MJD\,57163.68). Using the {\sc celldetect/CIAO}
procedure, the pulsar coordinates were determined as R.A.=
15$^{\rm h}$57$^{\rm m}$48.3$^{\rm s}$, Dec.=
-54$^\circ$24\arcmin53.1\arcsec\ (J2000) with $\simeq1$\arcsec\ uncertainty
(90\%). These coordinates differ by more than 6\arcsec\ from ones
mentioned in the {\it SIMBAD} database and are about 4.8\arcsec\ from those
reported by \citet{2008ATel.1371....1B}.

To clarify the situation and to understand the reasons for this difference,
we re-analysed both the archival {\it Swift}/XRT data (ObsID.\,00031096001 from
Jan 24, 2008, MJD\,54489.84) and the new data obtained during the current outburst on
Feb 6, 2015 (ObsID.\,00031096002, MJD\,57059.13). We calculated the enhanced source
position\footnote{http://www.swift.ac.uk/user\_objects/index.php}, where the
astrometry is derived using field stars in the UVOT images
\citep{goad2007,evans2009}. As a result, the best-fit source position was found as
R.A.= 15$^{\rm h}$57$^{\rm m}$48.35$^{\rm s}$, Dec.=-54$^\circ$24\arcmin52.7\arcsec\
(J2000, error radius 1.4\arcsec, 90\% confidence), which is in good agreement with the
{\it Chandra} measurements.

Thus we have determined, for the first time, accurate coordinates of \2s. These
coordinates were then used to search for its optical counterpart.

\begin{figure*}
\includegraphics[width=1.0\textwidth,bb=50 236 563 566,clip]{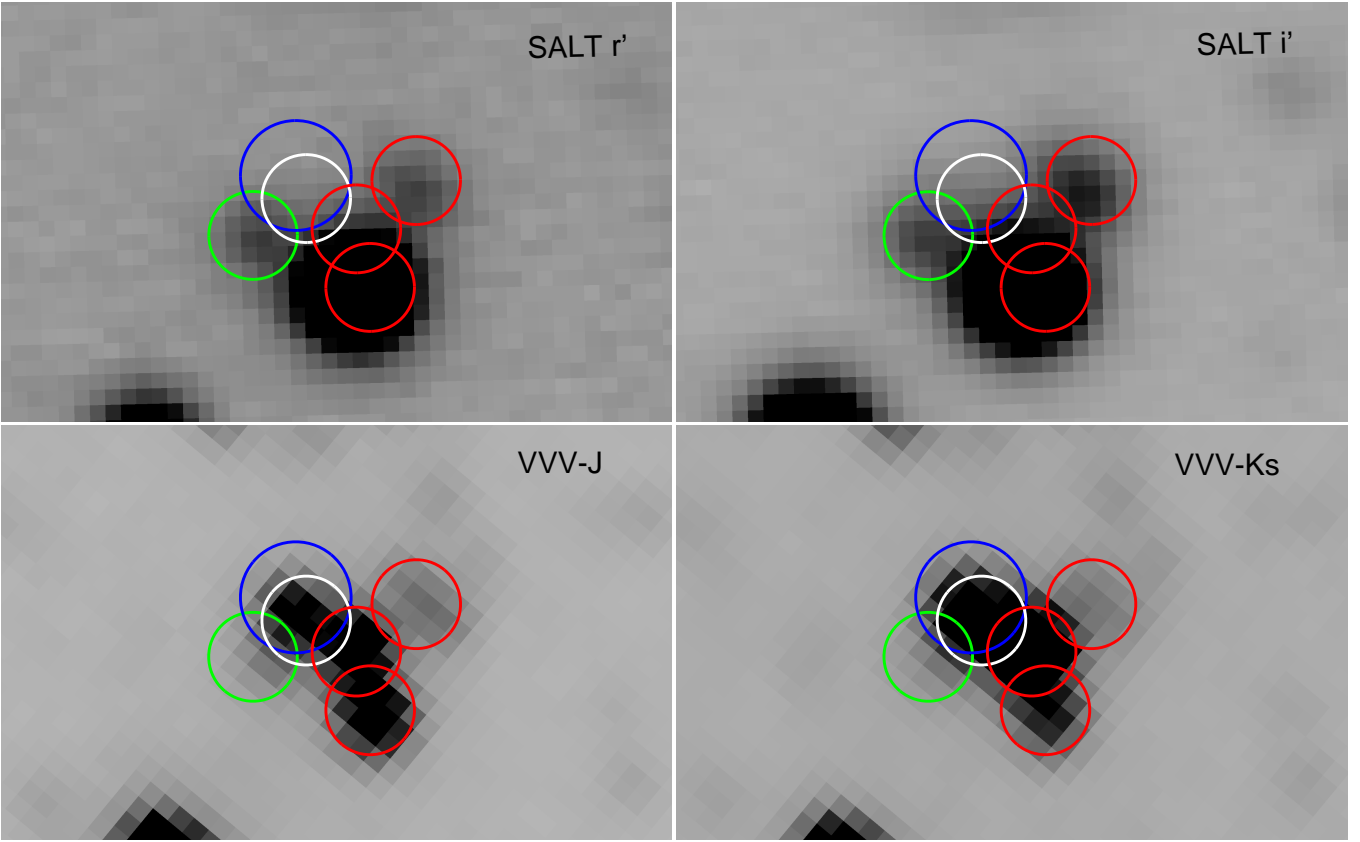}
\caption{SALTICAM $i'$ and $r'$ (upper panels) and \textit{VVV} $J$ and $Ks$
images of the region around \2s. The blue circle is the \textit{Chandra}
positional uncertainty (1\arcsec). The green circle is the position of the
SALTICAM star. The red circles denote nearby NIR stars, detected in the \textit{VVV}
images. The white circle is the position of the newly proposed counterpart of \2s.
The size of the circles around the optical and infrared stars is 0.8\arcsec\
and is just for illustrative purposes. Real positional uncertainties are
indicated in the text.}\label{image}
\end{figure*}

\section{Optical/IR Counterpart}

\subsection{{\it SALT} observations and data reduction}

In order to locate the optical counterpart of \2s, observations were taken of the
region indicated by the X-ray error circles from the \textit{Swift} and
\textit{Chandra} observations with the Southern African Large Telescope ({\it SALT})
\citep{Buckley2006} using the imaging camera, SALTICAM \citep{O'Donoghue2006}.
The first observation was taken on 11 Feb 2015 using $i'$ and
$H\alpha$ filters with 100 and 200\,s exposures, respectively.
This was followed by a spectroscopic observation on 22 Feb 2015 of the star $\sim4$\arcsec\ 
SE of the \textit{Swift} position, reported by
\citet{2008ATel.1371....1B} (the brightest star in the upper panels of
Fig.\,\ref{image}), using SALT/RSS \citep{Burgh2003, Buckley2008} with
the PG900 grating covering the range 3920$-$6990$\AA$, at an average spectral
resolution of 960, with a 1.5 arcsec slit. A single 1300\,s exposure revealed
a blue star with no emission features and no other compelling signatures
indicative of a HMXB optical counterpart. Following from the improved
\textit{Swift} position and the subsequent \textit{Chandra} observation, an
improved positional determination decreased the size of the error circle by a
factor of 3.5 in diameter and moved it $\simeq5$\arcsec\ to the east,
which then formally excluded the star previously observed.

A second, deeper, SALTICAM imaging observation was carried out on
15 Aug 2015 following the determination of the revised position. Five repeat
cycles of exposures in the $r', i'$ and $H\alpha$ filters, of 120\,s, 120\,s
and 600\,s, respectively, were obtained, plus an additional 120\,s $r'$-band
exposure. These observations were mostly taken in good seeing ($\sim$1
arcsec), with some degradation for the last few exposures, and in photometric
conditions. Fitted star image profiles typically had a FWHM in the range
1.2$-$1.4\arcsec. A faint star was detected near the edge of the new \textit{Chandra}
error circle, $\sim$2\arcsec\ NE of the previously observed star
(Fig.\,\ref{image}, green circle), which we term the SALTICAM star.

Photometry of star images in the vicinity of \2s was performed
in IRAF using standard routines in the DAOPHOT package. Given the reasonably crowded
field and the close proximity of the target to a brighter star, we used PSF fitting
routines instead of aperture photometry. After several trial fits, we found
that a Moffat function better modelled the stellar profiles and so was used
to produce the final instrumental magnitudes. Calibration of the instrumental
magnitudes was done using stellar catalogues
provided by the SuperCOSMOS Sky Survey \citep{Hambly2001}. The catalogue data
were positionally matched with the SALTICAM sources using a matched error of
1\arcsec. The $i'$ and $r'$ magnitudes of the resulting matches were plotted and
fit with a straight line function with gradient set to unity. The intercept
was used as a calibration zero-point and was added to the instrumental magnitude
of the target to produce the final calibrated magnitudes of the SALTICAM star, namely:

\medskip

$r' = 21.78 \pm 0.06$

$i' = 20.77 \pm 0.07$

\medskip

\noindent Note, that there is likely an additional systematic uncertainty of
about 0.1 mag due to the scatter in the data used to measure the zero-point offset.

To correct for interstellar extinction we adopt the standard $A_V - N_H$
relationship \citep{Predehl95} and the measured column density, namely
$N_{H}\simeq2.3\times10^{22}$ cm$^{-2}$ \citep[see Sect.4 and ][]{tsygankov2016}.
This gives an $A_V= 12.8$, and assuming the interstellar
absorption law given in \citet{Cardelli1989}, leads to $A_{r'}$ and $A_{i'}$ values of 11.3 and 8.7, respectively.

We then derive the following corrected magnitudes and colours for the SALTICAM
star:

\medskip

$r'_0 = 10.45 \pm 0.06 $

$i'_0 = 12.10 \pm 0.07$

$(r' - i')_0 = -1.01 \pm 0.09 $

$(R_C - I_C)_0 = -0.98 \pm 0.09 $

\medskip

The $(R_C - I_C)_0$ colour was derived using the relevant transformations from
the Sloan to the Johnson-Cousins systems \citep{Lupton2005}.
Note, that all BeXRBs have spectral types earlier than B2-B3
\citep{Negueruela1998, Negueruela2002}, with colours redder than the results of
our photometry. Together with the upper limits of the near IR magnitudes (see next section), this puts in doubt the connection between \2s\ and the SALTICAM star.

\subsection{Analysis of the {\it VVV} Survey data}

In order to further examine the validity of this potential counterpart to \2s
and to see if the SALTICAM star is also visible in the near infrared, we
downloaded the \textit{VVV} images for this region. These \textit{VVV} Survey
data were obtained in frames of the ESO programme 179.B-2002, with the VIRCAM
instrument, using $ZYJHKs$ filters (tile d\_061 with 80 s exposure for each filter).
Data were obtained thought the https://vvvsurvey.org service, which provides
access to reduced or fully calibrated data.

Surprisingly, we found there to be a total of 5 stars in the immediate vicinity
of the X-ray source position (Fig.\,\ref{image}, bottom panels) in the $JHKs$-bands
image (there are only 3 in the $i'$ image). To find the correct association
between the X-ray source and NIR star, photometry was performed in the same
way as described above for the SALTICAM images. The star from the
photometric catalogue with a position closest to the \textit{Chandra} position
was chosen as the most likely counterpart. Its coordinates are:
R.A.= 15$^{\rm h}$57$^{\rm m}$48.28$^{\rm s}$, Dec.=-54$^\circ$24\arcmin53.5\arcsec\
(J2000, error radius 0.07\arcsec, 90\% confidence), just
$\simeq0.5$\arcsec\ from the \textit{Chandra} coordinates. The calibrated $Z, Y,
J, H, K_s$ magnitudes for this star are:

\medskip

$Z = 18.59 \pm 0.12$

$Y = 17.18 \pm 0.09$

$J = 15.78 \pm 0.08$

$H = 14.46 \pm 0.14$

$K_s = 13.45 \pm 0.10$.

\medskip

Before constructing a Spectral Energy Distribution (SED) of the proposed counterpart,
we need to be sure that the star identified in the \textit{VVV} images is the most
probable counterpart to the X-ray source. On inspection of the images and positions
(see Fig.\ref{image}), we found that not only are the SALTICAM and \textit{VVV} sources
positionally inconsistent with each other (outside of the expected errors from our
astrometry -- the RMS from the SALTICAM analysis is 0.58\arcsec, averaged over both frames),
but the \textit{VVV} source is within 0.5\arcsec\ of the \textit{Chandra} position.
This, along with the $(R-I)$ colour derived above, suggests that the SALTICAM and
\textit{VVV} sources are in fact separate stars, and that the \textit{VVV} star is more
likely to be the correct counterpart. Further to this, we find that there is no source
detected in the \textit{VVV} image positionally consistent with the SALTICAM source,
and vice versa.

We then proceeded by treating the photometry of the two stars
separately in the subsequent analysis of the SEDs and distance determinations.

\subsection{SED and distance estimations}

To derive the SED for each star, we apply upper limits derived from each image.
Put simply, we assume that the optical star in the SALTICAM images is not
detected in the \textit{VVV} images, and thus we apply \textit{VVV} $ZYJHKs$
upper limits to this star.
These values were available in the headers of the downloaded \textit{VVV} images.
Likewise, we assume that the \textit{VVV} star within the {\it Chandra} error circle
was not detected in the SALTICAM images, and thus we apply formal $r'$ and $i'$
limits to this star. These were derived by plotting the brightness distribution of
all sources in each band, and checking where it turns over, and thus have a
statistical uncertainty, of up to $\pm$ 0.5 mags.

These limits were reddening corrected in the same manner as done previously
(as above, we used the \citet{Cardelli1989} formulae to determine the interstellar
absorption), resulting in interstellar reddening corrected SEDs for both stars.
All magnitudes (apparent, de-reddened, absolute for the distances 15 and 20 kpc)
as well as appropriate $A_\lambda$ values are presented in Table\,\ref{tabmag},
where (l) denotes an upper limit.

\begin{figure}
\includegraphics[width=\columnwidth,angle=0,bb=0 0 842 595,clip]{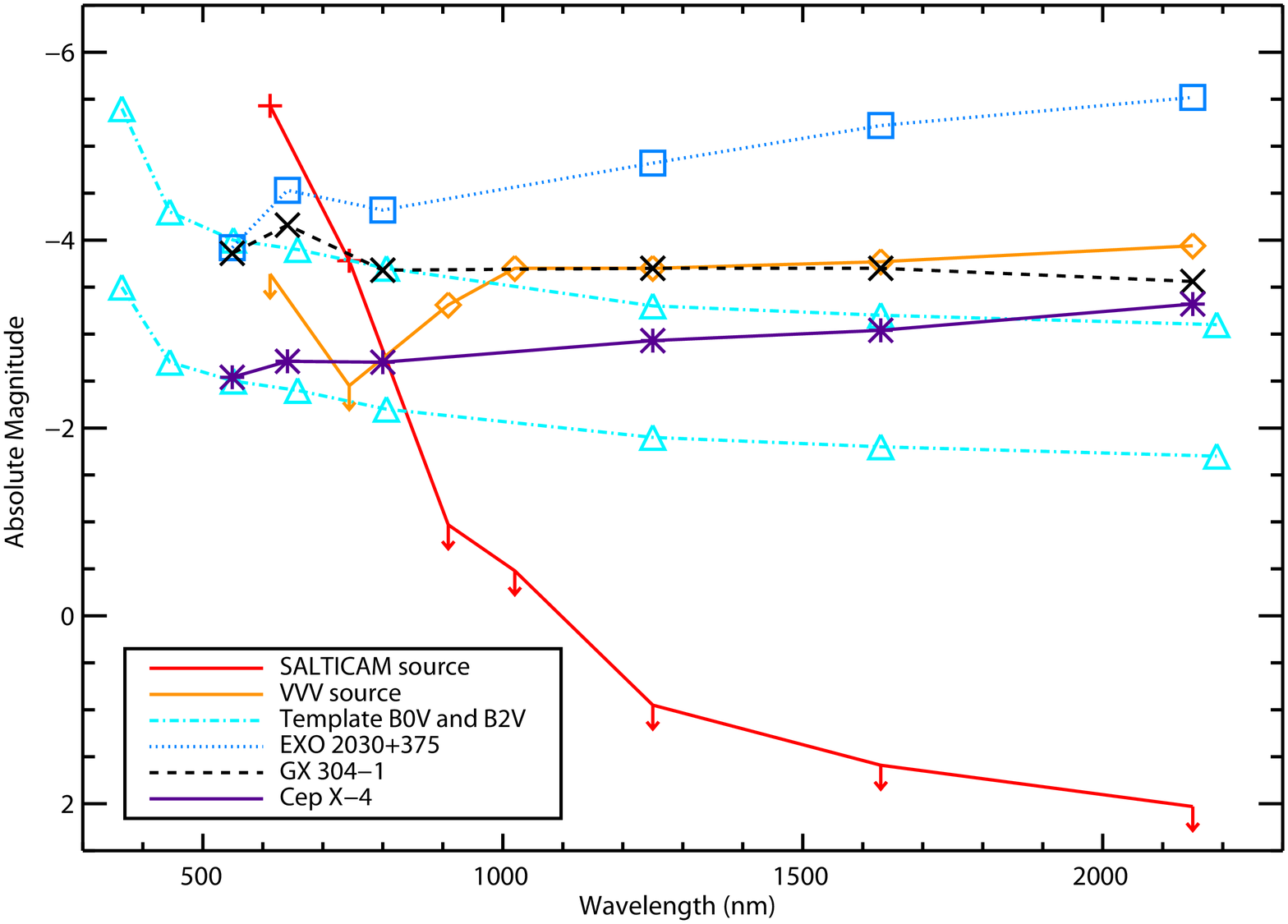}
\caption{The absolute magnitude of the SALTICAM (red) and \textit{VVV} (orange)
stars are plotted against wavelength to show their spectral energy distributions.
We also plot template SEDs for B0V and B2V (cyan) stars for comparison. The final
three curves represent the well known BeXRBs: EXO\,2030+375 (blue), GX\,304-1 (black)
and Cep\,X-4 (purple). The arrows denote the upper limits.}\label{sed}
\end{figure}

\begin{table*}
\centering
\caption{Magnitudes of the two potential optical counterparts of \2s\ described in the text}\label{tabmag}
\begin{tabular}{c|r|r|r|r|r|r|r|r|r}
  \hline
Band, & $A_{\lambda}$ & \multicolumn{4}{|c|}{SALTICAM star mags} & \multicolumn{4}{|c|}{\textit{VVV} star mags} \\[1mm]
Wavelength (nm) & & app & dered & 15 kpc & 20 kpc & app & dered & 15 kpc & 20 kpc\\
   \hline

$r'$, 612  & 11.34  & 21.78   & 10.45   & -5.44   & -6.06   & 23.6(l)& 12.26(l)& -3.62(l)& -4.24(l)\\
$i'$, 744  & 8.67   & 20.77   & 12.10   & -3.78   & -4.41   & 22.1(l)& 13.43(l)& -2.45(l)& -3.08(l)\\
$Z$, 909   & 6.02   & 20.93(l)& 14.91(l)& -0.97(l)& -1.60(l)& 18.59  & 12.57   & -3.31   & -3.94   \\
$Y$, 1020  & 5.00   & 20.40(l)& 15.40(l)& -0.48(l)& -1.11(l)& 17.18  & 12.18   & -3.70   & -4.33   \\
$J$, 1250  & 3.60   & 20.43(l)& 16.83(l)&  0.95(l)&  0.38(l)& 15.78  & 12.18   & -3.70   & -4.33   \\
$H$, 1630  & 2.35   & 19.82(l)& 17.47(l)&  1.59(l)&  0.96(l)& 14.46  & 12.11   & -3.77   & -4.40   \\
$K_s$, 2150& 1.51   & 19.43(l)& 17.92(l)&  2.04(l)&  1.42(l)& 13.45  & 11.94   & -3.94   & -4.56   \\

\hline
\end{tabular}
\end{table*}
It is immediately obvious that we have two very different potential counterparts.
To determine which SED is most like an early type Be star (as expected from
X-ray data), we converted the apparent magnitudes and limits above into
absolute magnitudes and plotted them against template SEDs for B0V and B2V
type stars. This result is presented in Fig.\ref{sed}. Note, that these
calculations were made under an assumption of a distance to the system
of 15 kpc. Such a large distance is expected from the previous analysis of
X-ray data of the source \citep{tsygankov2016}. Below we consider
this question based on the optical data. It is clearly seen from Fig.\ref{sed}
that the source detected in the \textit{VVV} images has an SED much more
similar in shape to an early type B star in comparison with the SALTICAM
star. Moreover, a moderate infrared excess likely due to a circumstellar
disk around the Be star is also apparent from the SED shape.

Confirmation of this hypothesis came after adding some known BeXRBs to
the figure. Photometry and distance estimates were found for EXO\,2030+375,
GX\,304-1 and Cep\,X-4 \citep{coe1997,riquelme2012,reig2014}. These were
converted to absolute magnitudes and plotted with the candidate counterparts in
Fig.\ref{sed}. One can see that all three BeXRBs are similar in shape
and magnitude to the \textit{VVV} source, lying between the B2 and B0
templates with moderate NIR excesses from the circumstellar material.

We looked at alternative interpretations for the {\it VVV} star in terms of
spectral type, distance and reddening, but no satisfactory alternative
could be determined. While it was possible to have a closer cooler star
with similar IR apparent magnitudes, the optical magnitudes of stars would
be many magnitudes brighter than the upper limits of the {\it VVV} star
imposed by the SALTICAM observations. Only for unrealistic combinations
of close distance ($< 500$ pc) and high reddening ($A_V > 12$) could we
approximate the observed magnitudes.

The approach we have made in trying to identify the likely nature of the
two possible optical counterparts is as follows. Firstly, we have constructed
a spreadsheet consisting of the absolute magnitudes of stars of varying
spectral type and luminosity class considered appropriate either as the bona
fide optical counterpart of \2s\ or a potential field star, which simply
happen to be inside the error circle, unassociated with the X-ray source.
For the former we considered stars of spectral type B2V to O8V, while for
the latter we considered stars of a range of spectral types, including M
dwarfs and white dwarfs, as potential field stars.

We derived the appropriate $A_\lambda/A_V$ values for each of the bandpasses
($r', i', Z, Y, H, J, K_s$) using the formulae in \citet{Cardelli1989}. These empirical
relations are somewhat of a simplification for the true interstellar extinction,
but for the purposes of estimating SEDs, we consider they are adequate.
With the $A_V$ determined from $N_H$ \citep{Predehl95}, the apparent
magnitudes were then determined for a given assumed distance. For a star
considered as a potential optical counterpart to \2s, we fixed the observed
$N_H$ value derived earlier and adjusted the distance such that the
calculated apparent (reddened) magnitude matched the measured magnitudes

In the case of a field star unrelated to \2s, we allowed both $N_H$ and
distance to be variables to investigate if we could obtain self consistent
magnitudes and colours.

As can be seen in Fig.\,\ref{sed}, the SALTICAM star is too blue to be
consistent with a Be star of spectral type earlier than B2V, which is why
we discount it as the likely optical counterpart of \2s. We investigated
the potential nature of the star by looking at the absolute magnitudes
and colours of various luminosity class V stars \citep{pec13} and white
dwarfs, comparing them to the observed $r'$ and $i'$ magnitudes and the
upper limits of the $JHK$ magnitudes from the {\it VVV} survey.  The
conclusion from this investigation was that no class V star is able to
match the observations. Only for spectral types earlier than G0 are the
intrinsic colours consistent with the observations, but the apparent
magnitudes would place such a star at a distance of $\sim$20 kpc and the
interstellar reddening would make the star relatively brighter at IR
magnitudes, inconsistent with the {\it VVV} upper limits. Cooler spectral
types (K or M) would reduce the distance, but even for unreddened M-dwarfs,
their colours are much too red to be consistent with the observations.
The only possibility we could find to explain the observations is if the
SALTICAM star is a nearby ($\sim$ 100 pc) cool white dwarf, whose (unreddened)
magnitude and colours \citep{Chabrier2000} would be consistent with the
observations.

A precise distance estimate to the source is difficult, given only optical
limits are placed by the SALTICAM photometry of the {\textit VVV} star.
Figure \ref{sed} shows where the SED of the source would sit against the
absolute magnitudes of B0V and B2V template stars assuming a real distance of
15 kpc. Assuming that the most likely spectral type of the source is between
B0-B2 (given the BeX nature of the source), we expect a distance of less than
15 kpc to be unlikely. Conversely, placing the source SED onto
Fig.\,\ref{sed} after assuming a real distance of 20 kpc (see
Table\,\ref{tabmag}) is consistent with a B1V star, but is still lower than a
B0V star. Thus, we also tentatively suggest that a B0V spectral type is
unlikely as the distance would put the source outside of the Galaxy.

The galactic coordinates of \2s are $l = 327.94^{\circ}$ and  $b = -0.86^{\circ}$,
which places \2s\ on the other side of the Galaxy in the Scutum-Centaurus arm or
even further in the Sagittarius arm \citep{urquard2014}.

\section{X-ray spectroscopy}

\begin{table*}
\centering
\caption{Best-fitting spectral parameters for the three observations of \2s\ described in the text.}\label{tabspec}
\begin{tabular}{llll}
\hline
\hline
Parameter &  \multicolumn{3}{c}{Instrument, MJD}\\[1mm]
\hline
          &  XRT   & XRT   & Chandra  \\[1mm]
          &  54489.84 &  57059.13 & 57163.68 \\

\hline
$N_H$, $10^{22}$      & $2.42\pm0.23$ & $2.38\pm0.33$ & $1.7$ (fix) \\
$kT_{\rm BB}$, keV    & $2.27\pm0.12$ & $1.79\pm0.13$ & $1.4^{+0.3}_{-0.2}$ \\
$norm_{\rm BB}$       & $6.32\pm0.94$ & $8.85\pm2.11$ & $1.09^{+0.66}_{-0.44}\times10^{-2}$ \\
$\chi^2$ (d.o.f)      & $1.01 (117)$  & $0.72 (39)$   & $0.90 (37)$ \\ [2mm]
\hline
$N_H$, $10^{22}$      & $3.93\pm0.36$ & $4.20\pm0.54$ & $1.7$ (fix)\\
Photon index          & $0.65\pm0.11$ & $1.16\pm0.18$ & $0.86^{+0.27}_{-0.27}$ \\
$\chi^2$ (d.o.f)      & $1.08 (117)$  & $0.78 (39)$   & $0.91(37)$     \\ [2mm]
\hline
Flux (0.5-10 keV)     & $(1.06\pm0.04)$ & $(0.68\pm0.06)$ & $3.30^{+0.23}_{-1.36}$ \\
    \flux             & $\times10^{-9}$ & $\times10^{-9}$ & $\times10^{-13}$ \\

\hline
\end{tabular}
\end{table*}
The detailed analysis of the broadband (3-79 keV) spectrum of \2s\ was
recently performed by \citet{tsygankov2016} using data from the {\it NuSTAR}
observatory. It was shown that the spectral continuum contains two components
-- a thermal black body emission at low energies and a power law with an
exponential cutoff at higher energies. Additionally, a cyclotron absorption
line at an energy of $\simeq23.5$ keV and a fluorescent iron emission line at
6.4 keV were detected \citep{tsygankov2016}.

The working energy bands of {\it Swift/XRT} and {\it Chandra} have an upper
bound near $\sim10$ keV, and the effective area of both of them are
significantly reduced at energies above 7-8
keV\footnote{http://swift.gsfc.nasa.gov/about\_swift/xrt\_desc.htm}\footnote{http://cxc.harvard.edu/proposer/POG/html/chap6.html}.
Therefore, we found it unnecessary to use such a complex model to fit our
data.

\begin{figure}
\centering
\includegraphics[width=0.98\columnwidth,bb=45 239 546 689,clip]{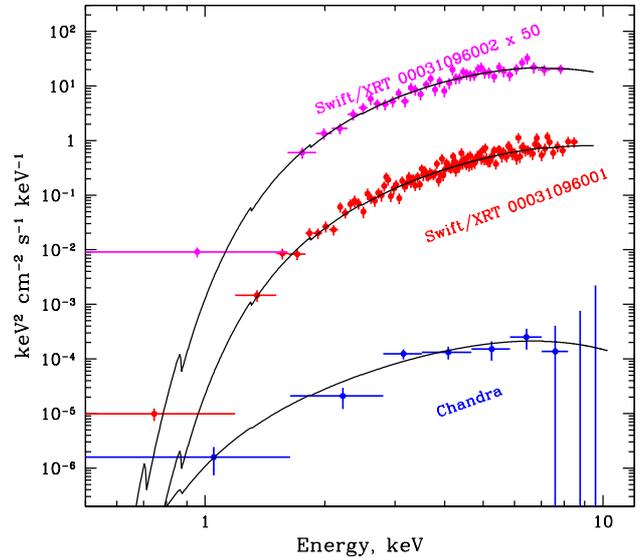}
\caption{Spectra of \2s, obtained with the {\it Swift}/XRT (magenta and red circles)
and {\it Chandra} observatory (blue circles). The spectrum of the second XRT
observation is multiplied by a factor of 50 for clarity. Solid lines represent
the best-fit models (see text for details).}\label{xspec}

\end{figure}

We found that all three source spectra in the 0.5-10 keV energy band can be
well approximated by either absorbed black body emission ({\sc phabs*bbodyrad}
model in the {\sc XSPEC} package) or an absorbed power law ({\sc phabs*powerlaw}
model in the same package). The corresponding best-fit parameters are presented in
Table\,\ref{tabspec} together with measured fluxes and $\chi^2$ values. Note that the
former model describes the spectra slightly better. This is due to the fact that
in the spectrum of \2s\ there is a relatively low exponential cutoff at $\sim5.5$ keV
\citep{tsygankov2016}. The inclusion of this parameter in the model formally
improves the fit and reduces the absorption value, but does not provide a
better determination for the parameters due to the limited energy band of the instruments.
It is shown in Table \ref{tabspec} that the spectral parameters (e.g., temperature
and absorption) are quite similar for all observations. The measured absorption
value in the high state is in good agreement with the \textit{NuSTAR} results
\citep{tsygankov2016} and only slightly higher than the one of the galactic
interstellar absorption in the direction of \2s,
$N_{H, Gal}\simeq1.7\times10^{22}$ cm$^{-2}$ \citep{LAB}. The absorption value
in the low state is determined poorly due to insufficient statistics of the
{\it Chandra} data. The value varied from $0.7\times10^{22}$ to
$1.5\times10^{22}$ cm$^{-2}$ depending on the spectral model, but due to
large uncertainties it agrees with the interstellar value at the $2\sigma$ level.
Therefore, we fixed $N_{H}$ to the interstellar value
and fitted only the temperature or power law index and corresponding normalization.
Note, that the fixing of $N_{H}$ at $2.3\times10^{22}$ cm$^{-2}$ practically
does not significantly change results of the approximation.

The inclusion of an iron emission line didn't improve the goodness of the
fit for the first XRT observation and only improved it insignificantly
for the second. Therefore, we obtain only upper limits ($1\sigma$) for
the iron line equivalent width in these observations: 116 and 390 eV,
respectively (the line energy and width were fixed at 6.4 and 0.1 keV).
X-ray spectra of \2s\ are shown in Fig.\ref{xspec} along with best-fit models
in the form of the absorbed black body emission.

Of particular interest is the observation with the {\it Chandra} observatory, during which the
source flux was almost 4 orders of magnitude lower than that near the peak
of the outburst. Observations of very faint X-ray transients can be used
for testing different models of the neutron star cooling \citep[see e.g.][]{wijnands2013}
and its emission in the quiescent state. In particular, based on the spectral
shape, it can be determined whether the source is still accreting at low level or
whether the heated neutron star is observed. Unfortunately, the statistics of the {\it Chandra}
observation are not sufficient to distinguish between thermal or non-thermal
emission of \2s\ in the low state (Table\,\ref{tabspec}). Additionally, the
temporal resolution of the ACIS instrument (about 3.2 sec) makes detection of
the 9.3\,s spin period in \2s\ very difficult, particularly with so few counts.

\section{Conclusions}

In this paper we have used {\it Chandra} and {\it Swift}/XRT data to
accurately determine the coordinates of the long known, but
poorly studied, transient X-ray pulsar \2s\ for the first time.

The subsequent optical observations with the \textit{SALT} telescope
and a comprehensive analysis of the \textit{VVV} survey data led to
the identification of its likely infrared counterpart, whose spectral
energy distribution is consistent with a Be-star, as expected from its
previous classification as a BeXRB with an accreting X-ray pulsar.
Moreover, a comparison of the measured SED with those of known Be/X-ray
binaries demonstrates the latter are similar and has allowed us to estimate
the spectral type of the \2s\ companion
star as B1-2V and the distance to the system as $>15$ kpc. This value
agrees well with the independent estimations of $d=20\pm4$ kpc from the temporal
properties of the pulsar \citep{tsygankov2016}.

These distance estimates place \2s\ in the Scutum-Centaurus arm, on the far
side of the Galaxy from the Sun, or even further, depending on the possible
spectral type of its companion star. This result can be
considered as the first reliable distance determination to \2s\ and, for an
outlying Galactic object, makes \2s\ one of the most distant Galactic
high-mass X-ray binaries known. We note that estimations of distances to
other distant Galactic sources have, as a rule, quite large uncertainties and
still require additional confirmations \citep[see,
e.g.,][]{tsygankov2005,shaw09,karas10,pel11}. In turn, the determination of
distances to high-mass X-ray binaries at the furthest regions of the Galaxy,
behind the Galactic Center, is very important for determining the space
density of such objects \citep{2013MNRAS.431..327L}.

The measurement of the distance to \2s\ ($d>15$ kpc) has allowed us to
estimate a lower limit to its unabsorbed luminosity near the outburst maximum
as $L_{0.5-10 keV}\simeq3.2\times10^{37}$ \lum\ and in the faint state
$L_{0.5-10 keV}\simeq1.1\times10^{34}$ \lum, from the {\it Swift}/XRT and
{\it Chandra} data, respectively. Taking into account that the ratio between the
X-ray bolometric source flux and its flux in the 0.5-10 keV energy band is
about 2 (from the analysis of the source broadband spectrum, obtained by {\it
NuSTAR} and our results, which demonstrate that its spectral shape is not
dependent on the luminosity) we can estimate a lower limit to the maximum
(bolometric) X-ray luminosity of the source $L_{X, max}\simeq6.4\times10^{37}$ \lum.
The value $L_{X, max}\simeq1.2\times10^{38}$ \lum, obtained for a distance of 20 kpc,
is comparable with that measured for other bright Be/X-ray binary transients
such as V\,0332+53 and 4U\,0115+63, and implies that the observed outbursts
from \2s\ are Type II in nature. The corresponding source luminosity in quiescence, namely
$L_{X, faint}\sim(2.2-4)\times10^{34}$ \lum\, is high enough for testing different
models for the low accretion regime and neutron star cooling scenario, should
one obtain sufficiently long exposures in the faint X-ray state.

\section*{Acknowledgments}

This research has made use of {\it Chandra} data and software provided by the Chandra X-ray
Center in the application package CIAO. We also used of data supplied
by the UK Swift Science Data Centre at the University of Leicester. The optical
observations were undertaken with SALT Directors Discretionary Time under programs
2014-2-DDT-004, 2014-2-DDT-005 and 2015-1-DDT-004, for which we are grateful.
The obtained results are also based on data products from \textit{VVV} Survey observations
made with the {\it VISTA} telescope at the ESO Paranal Observatory under programme ID 179.B-2002.
AL, DB acknowledge support by the bilateral Russian-South African grant RFBR 14-02-93965.
AL acknowledges support from the Dynasty Foundation and grant NSh-6137.2014.2.
DB acknowledges support through the National Research Foundation of South Africa.
LJT acknowledges support from the Claude Leon Foundation. We also especially thank
to Ignacio Negueruela, who was a referee of this paper, for the very important comments
and suggestions to use the VVV data. LJT would like to thank Petri Vaisanen for
useful discussions on SALTICAM calibrations and photometric limits.

\bsp    
\label{lastpage}
\end{document}